%% file: B2G-12-022_temp.tex
\pdfoutput=1

\documentclass[11pt,twoside,a4paper,cmspaper,final,collab]{cms-tdr}
\def\svnVersion{280970}\def\cmsCernNo{2013-037}\def\cmsCernDate{\today}\def\cmsMessage{Published in Physical Review Letters as \href{http://dx.doi.org/10.1103/PhysRevLett.114.101801}{\doi{10.1103/PhysRevLett.114.101801.}}}

\begin{document}\cmsNoteHeader{B2G-12-022}
\typeout{**** Begin including local definitons.tex file *****}
\typeout{**** tdr loads this quietly without a specific input command *****}

\newcommand{\fix}[1]{{\bf <<< #1 !!!} }
 
\typeout{**** End including local definitons.tex file *****}

\def\roostats{\texttt{RooStats}\xspace}
\def\L{{\ensuremath{\cal L}}\xspace}
\def\bnu    {\ensuremath{{\boldsymbol\nu}}\xspace}
\def\bsigma {\ensuremath{{\boldsymbol\sigma}}\xspace}
\def\selEff{\ensuremath{\epsilon_\mathtt{sel}}\xspace}
\def\acc{\ensuremath{{\cal A}}\xspace}

\def\proton      {\ensuremath{p}\xspace}

\hyphenation{had-ron-i-za-tion}
\hyphenation{cal-or-i-me-ter}
\hyphenation{de-vices}

\RCS$Revision: 280963 $
\RCS$HeadURL: svn+ssh://svn.cern.ch/reps/tdr2/papers/B2G-12-022/trunk/B2G-12-022.tex $
\RCS$Id: B2G-12-022.tex 280963 2015-03-17 15:10:10Z kkotov $
\newlength\cmsFigWidth
\ifthenelse{\boolean{cms@external}}{\setlength\cmsFigWidth{0.95\columnwidth}}{\setlength\cmsFigWidth{0.65\textwidth}}
\ifthenelse{\boolean{cms@external}}{\providecommand{\cmsLeft}{top}}{\providecommand{\cmsLeft}{left}}
\ifthenelse{\boolean{cms@external}}{\providecommand{\cmsRight}{bottom}}{\providecommand{\cmsRight}{right}}
\ifthenelse{\boolean{cms@external}}{\providecommand{\CL}{C.L.\xspace}}{\providecommand{\CL}{CL\xspace}}
\newcommand{\CLs}{\ensuremath{\mathrm{CL}_\mathrm{s}}\xspace}
\cmsNoteHeader{B2G-12-022} 
\title{Search for Monotop Signatures in Proton-Proton Collisions at \texorpdfstring{$\sqrt{s}=8\TeV$}{sqrt(s) = 8 TeV}}

\date{\today}

\abstract{
Results are presented from a search for new decaying massive particles whose
presence is inferred from an imbalance in transverse momentum and which are
produced in association with a single top
quark that decays into a bottom quark and two light quarks.
The measurement is performed using 19.7\fbinv of data from proton-proton collisions at a center-of-mass energy of 8\TeV,
collected with the CMS detector at the CERN LHC.
No deviations from the standard model predictions are observed
and lower limits are set on the masses of
new invisible bosons. In particular, scalar and vector particles, with masses below 330 and 650\GeV, respectively, are excluded at 95\%
confidence level,
thereby substantially extending a previous limit published by the CDF Collaboration.
}

\hypersetup{%
pdfauthor={CMS Collaboration},%
pdftitle={
Search for monotop signatures in proton-proton collisions at sqrt(s) = 8 TeV},
pdfsubject={CMS},%
pdfkeywords={CMS, physics, top quark, monotop}}

\maketitle 
Extensions of theories beyond the standard model (BSM),
such as those with universal extra dimensions~\cite{Appelquist:2000nn}
or supersymmetry~\cite{Nilles:1983ge,Haber:1984rc},
predict the existence of neutral massive particles that are ``invisible'',
that is, they interact only weakly with matter.
Such particles can be produced in collider experiments, but escape detection so that their
existence can only be inferred by the presence of
a large imbalance in transverse momentum (\MET). Both the ATLAS~\cite{Aad:2008zzm}
and CMS~\cite{Chatrchyan:2008aa} collaborations have performed searches
for the invisible BSM particles in monojet~\cite{ATLAS:2012ky,Khachatryan:2014rra} and 
monophoton~\cite{Aad:2012fw,Chatrchyan:2012tea} signatures that manifest themselves through the presence
of a single jet or photon associated with large \MET. These searches have not revealed any evidence for
BSM monojet or monophoton final states, but this nonobservation
can be accommodated in a theory where the new particles
change the quark flavor and convert a light quark to a top quark.
In this case, the BSM event signature would not correspond to monojet or monophoton final states, but to events containing single top quarks and large \MET, referred to as ``monotop'' candidates. In this Letter, we present a search for such events in which an invisible BSM particle is produced in association with a top quark~\cite{Andrea:2011ws,Agram:2013wda,delAguila:1999ac,Morrissey:2005uza,Desai:2010sq,Dong:2011rh,Davoudiasl:2011fj,Wang:2011uxa,Kamenik:2011nb,Alvarez:2013jqa,Kumar:2013jgb}.

Depending on the spin statistics of the invisible particle, at tree level a monotop system can be produced through two main mechanisms: it can originate either (i)
from the decay of a heavy bosonic resonance, with \MET arising from an invisible baryon-number violating fermionic state
(for instance, $\cPaqd\cPaqs\to\sUp_i\to\cPqt\PSGczDo$, where \cPaqd~and \cPaqs~denote anti-\cPqd~and anti-\cPqs~quarks,
$\sUp_i$ are any of the up-type squarks of $R$-parity violating supersymmetry~\cite{Barbier:2004ez}, and \cPqt~and \PSGczDo~are the top quark and neutralino),
or (ii) through flavor-changing (FC) interactions mediated by an invisible bosonic state (for instance, $\cPqu\cPg\to\cPqu\to\cPqt\upsilon$,
where \cPqu, \cPg, \cPqt, and $\upsilon$ are an up quark, gluon, top quark, and invisible BSM particle, respectively)~\cite{Andrea:2011ws}.
In both cases, the new invisible particles are assumed to have a branching fraction
close to unity for decay to hidden sector particles.
While this requires some degree of tuning, such scenarios are well motivated as discussed in Ref.~\cite{Boucheneb:2014wza}.
Consequently, even in the presence of nonvanishing couplings to
SM particles, which is necessary for the production of such invisible particles in collider experiments,
an \MET signature is expected. The present study focuses on the second class of the above-mentioned processes where a bosonic invisible state is produced that yields a large \MET in association with a single top quark that
decays to a bottom quark and a \PW~boson, with the latter decaying into a pair of quarks.

The search is performed on data from proton-proton collisions recorded at a center-of-mass energy of 8\TeV, corresponding to an
integrated luminosity of 19.7\fbinv, and recorded with the CMS detector~\cite{Chatrchyan:2008aa} at the CERN LHC.
The most important backgrounds for the event signature with three jets and large \MET are \cPZ+jets, \PW+jets, and \ttbar processes.
The \cPZ+jets and \PW+jets backgrounds are estimated from data, and the signal yield is determined simultaneously with multijet background,
using a likelihood approach based on the observed multiplicity of \cPqb-tagged jets.

\begin{figure}[htbp]
\begin{center}
\includegraphics[width=0.22\textwidth]{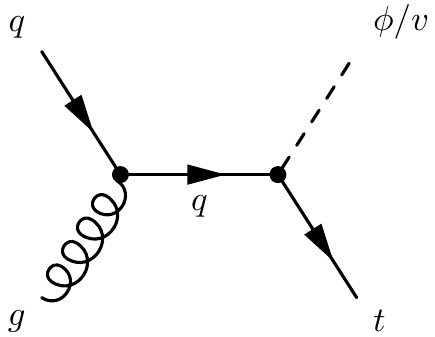}
\includegraphics[width=0.22\textwidth]{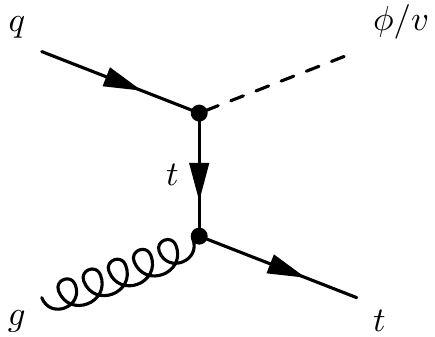}
\caption{Feynman diagrams for $s$-channel (left) and $t$-channel (right) monotop production.}
\label{fig:diagrams}
\end{center}
\end{figure}

We interpret the results within a simplified field theory~\cite{Andrea:2011ws,Agram:2013wda}
where the invisible particle can be either a scalar ($\phi$) or a vector ($v$) boson, with its Lagrangian given by
\begin{linenomath}
\begin{equation}
\mathcal{L} = \mathcal{L}_\mathrm{SM} + \mathcal{L}_\text{kin} + a^\mathrm{0}_\mathrm{FC}\phi\bar{u}u + a^\mathrm{1}_\mathrm{FC}v_\mu\bar{u}\gamma^\mu u + \text{H.c.},
\end{equation}
\end{linenomath}
where $\mathcal{L}_\mathrm{SM}$ denotes the SM Lagrangian, $\mathcal{L}_\text{kin}$
kinetic terms for the $\phi$ and $v$ fields, and the remaining terms model the interactions
of the invisible states with up-type quarks. The coupling strengths are
embedded in two $3\times 3$ matrices in flavor space ($a^{0,1}_\mathrm{FC}$), where only the elements
connecting the first and third generations are nonvanishing and set to 0.1, as done in Ref.~\cite{Andrea:2011ws}.
The Feynman diagrams for tree-level production are shown in Fig.~\ref{fig:diagrams}.

Events are recorded using a trigger requiring $\MET>150\GeV$.
For the background estimation we use an independent control sample of events
with an isolated single-muon trigger with a transverse momentum threshold of $\pt>24\GeV$.

The monotop model is implemented within the \textsc{FeynRules}
package~\cite{Christensen:2008py,Alloul:2013bka} and is interfaced~\cite{Christensen:2009jx,Degrande:2011ua} to the \MADGRAPH~5 event
generator~\cite{Alwall:2014hca}. Simulated events are produced for masses of invisible particles from 0 to 0.2\TeV in steps of 0.05\TeV, and from 0.2 to 1\TeV in steps of 0.1\TeV.
The production cross sections are calculated at leading order (LO) with \MADGRAPH using CTEQ6.1L~\cite{Pumplin:2002vw} parton distribution functions (PDF).
This PDF set is also used for simulating the SM background processes.

The main backgrounds are generated using \MADGRAPH. For \PW+jets and \cPZ+jets processes,
events are generated including up to four additional partons.
For \cPZ($\to\Pe\Pe,\Pgm\Pgm,\Pgt\Pgt$)+jets (with mass $m_{\ell\ell}>50$\GeV) and $\PW(\to\Pe\PGn,\Pgm\PGn,\Pgt\PGn)+\text{jets}$ processes we use
the next-to-next-to-leading order (NNLO) cross sections of
3.50\unit{nb} and 37.5\unit{nb}, respectively, as calculated using the {\FEWZ}~3.1 program~\cite{Gavin:2010az}.
For the $\cPZ(\to\PGn\PGn)+\text{jets}$ process we use events generated in four bins
of \HT, the scalar sum of \pt of all of the generated partons in the process: $50<\HT<100$\GeV,
$100<\HT<200$\GeV, $200<\HT<400$\GeV, and $\HT>400$\GeV, with respective
LO cross sections of 381, 160, 41.5, and 5.27\unit{pb}.
The \ttbar sample includes up to three additional partons at the matrix element level,
and is rescaled to an inclusive NNLO cross section of 246\unit{pb}~\cite{Czakon:2013goa}.

Other SM backgrounds arise from single top quark and diboson ($\PW\PW$, $\PW\cPZ$, and $\cPZ\cPZ$) production.
Single top quark production is modeled with \POWHEG~1.0~\cite{Nason:2004rx,Frixione:2007vw,Alioli:2009je,Alioli:2010xd}, and diboson production
is modeled with \PYTHIA~6.4.22~\cite{Sjostrand:2006za}. All Monte Carlo (MC) generated events are evolved using \PYTHIA~6.4.22 with Z2$^\ast$ tune~\cite{Field:2011iq}
and processed with a full simulation of the CMS detector implemented in the \GEANTfour
package~\cite{Agostinelli:2002hh}.

We require at least one reconstructed primary vertex and reject events with evidence of significant beam halo
or events with a large amount of detector noise~\cite{Chatrchyan:2009hy}.

The CMS particle-flow (PF) algorithm~\cite{CMS:2009nxa,CMS:2010byl,CMS:2010eua} is used to reconstruct
and identify each particle with an optimized combination of information from all the CMS subdetectors.
The only charged PF particles considered in reconstructing an event are those associated with the main primary vertex,
which is defined as the primary vertex with the largest sum of $\pt^2$ of all the associated tracks.
Particles identified as originating from other collisions in the beam crossing (pileup)
are removed from consideration.

The three jets from $\cPqt\to\cPqb\PW\to\cPqb\cPq\cPaq'$
decay are reconstructed using the anti-\kt clustering algorithm~\cite{Cacciari:2008gp} with a distance parameter of 0.5.
During jet reconstruction, the charged particles arising from pileup interactions are excluded,
while the neutral pileup component is accounted for using the area-based energy subtraction procedure described in Refs.~\cite{fastjet1,fastjet2}.
Jet energy corrections used in this measurement rely on simulation and on studies performed in data~\cite{Chatrchyan:2011ds}. Only jets with $\pt\geq35\GeV$ and $\abs{\eta}<$~2.4 are considered,
where $\eta$ is the pseudorapidity.
The two highest-\pt (leading) jets must have a $\pt>60\GeV$, while the \pt of the jet with third highest \pt
has to be above 40\GeV.
The invariant mass of the three jets has to be less than 250\GeV. Events containing additional jets with a \pt above 35\GeV are rejected.
We require one of the three jets to be identified as a candidate jet from a \cPqb~quark.
The identification algorithm is based on the reconstruction of a displaced secondary vertex~\cite{Chatrchyan:2012jua}.
A jet from a \cPqb~quark is tagged with $\approx$70\% efficiency. The probability to tag a light jet from
\cPqu, \cPqd, or \cPqs~quarks, or a gluon jet is 1--4\% depending on the jet \pt.
We also use events without \cPqb-tagged jets in order to define a background-enriched sample to extract the
normalization for multijet background.

Signal-candidate events containing isolated muons or electrons are rejected.
Muons are reconstructed by matching tracks from the outer muon detector to tracks reconstructed by the inner tracker~\cite{Chatrchyan:2013sba}.
Muons are required to have $\pt>10\GeV$ and $\abs{\eta}<2.4$.
Electrons are reconstructed by associating tracks from the inner tracker to clustered energy depositions in the electromagnetic calorimeter~\cite{CMS-PAS-EGM-10-004}.
Electrons are required to have $\pt>20\GeV$ and be within $\abs{\eta}<2.5$, excluding the transition
region between barrel and endcap defined by $1.44<\abs{\eta}<1.57$.
Standard CMS muon and electron identification criteria~\cite{Chatrchyan:2013sba,CMS-PAS-EGM-10-004} are applied.
The muon (electron) relative isolation variable $I_{\text{rel}}$ is computed by first summing
the transverse momenta of the reconstructed particles in a cone of $\Delta R=\sqrt{\smash[b]{(\Delta \eta)^2 + (\Delta \phi)^2}}<0.4\,(0.3)$ around the muon (electron) direction,
excluding the contribution of the lepton, and then dividing this sum by the transverse momentum of the lepton.
The lepton candidates are rejected if they satisfy $I_{\text{rel}}<0.2$.

The \MET vector is defined by the negative vector
sum of the transverse momenta of all the reconstructed particles in the event.
The \MET threshold of 350\GeV used in the analysis is optimized to give the
most sensitive exclusion limit on the production cross section.
This threshold is also nearly optimal for attaining best significance in the signal.

The dominant backgrounds after implementing the selection criteria are \ttbar and $V$+jets events, with $V$ being either a \cPZ~or a \PW~boson.
For electroweak vector boson production up to three additional jets are considered, leading to a large systematic uncertainty in the predicted production rate.
For this reason, we estimate the $V$+jets background using data.

The control region for \PW+jets and \cPZ+jets backgrounds is
defined with an alternative selection,
requiring one or two isolated muons in addition to the three jets.
A tighter selection is applied for muons, requiring them
to satisfy $\pt\geq40\GeV$ and $\abs{\eta}<2.1$. In this case, the relative combined isolation variable in
a cone of $\Delta R<0.4$ must be below 0.12. As in the signal selection, the three jets are required to
have $\pt> 60$, 60, 40\GeV respectively, and the invariant mass of the three jets has to be less than 250\GeV.
Events with any additional jets with a \pt above 35\GeV, as well as events with additional isolated electrons or
muons are rejected.

The $\cPZ(\to\PGn\PGn)$+jets background is estimated from events with two muons and three jets.
In such events, we replace the requirement for $\MET>350\GeV$ with the requirement for the vector sum of
the \pt of the two muons and of \MET to be greater than 350\GeV.
We also suppress the non-\cPZ~backgrounds by selecting events with \MM invariant mass between 60 and 120\GeV.
The residual non-\cPZ~backgrounds are reduced to 1.5\%; thus they make a negligible contribution to the overall uncertainty.
The $\cPZ(\to\Pgn\Pgn)$+jets background is calculated using the following equation:
\begin{linenomath}
\begin{equation}
\label{eq:ddDYbg}
N(\cPZ\to\Pgn\Pgn) = \frac{ N^{\text{obs}}(\Pgm\Pgm) }{ A\times\epsilon(\Pgm\Pgm) }
 \frac{ \mathcal{B}(\cPZ\to\Pgn\Pgn) }{ \mathcal{B}(\cPZ\to\Pgm\Pgm) }
\end{equation}
\end{linenomath}
where $N^{\text{obs}}(\Pgm\Pgm)$ is the number of observed events with two muons, $A\times\epsilon(\Pgm\Pgm)$ is the product of
acceptance and efficiency to identify and select the two muons, as measured in simulation, and $\mathcal{B}(\cPZ\to\Pgn\Pgn)/\mathcal{B}(\cPZ\to \Pgm\Pgm) = 5.94$~\cite{Beringer:1900zz}
is the ratio of branching fractions for \cPZ\ decays into two neutrinos and two muons.
The accuracy of the background estimate is limited by the number of selected events with two muons.
The estimated $\cPZ(\to\Pgn\Pgn)$+jets background is presented in Table~\ref{tab:mc2data}.

The $\PW(\to\ell\Pgn)$+jets background is calculated from events with a single muon and three jets.
Just as in the selected signal events, \MET has to be greater than 350\GeV. The transverse mass constructed with the muon-\pt and \MET vectors has to be less than 180\GeV.
From simulation we estimate the single-muon background that does not arise from \PW\ boson production (roughly a third of events), and subtract it from the observed
number of events. The resulting number is divided by the acceptance and efficiency of the single-muon
selection, providing thereby the number of $\PW(\to\Pgm\Pgn)$+jets events.
Assuming lepton universality, we use the same estimate for events with other lepton flavors.
In the simulation, we calculate the probability that the $\PW(\to\ell\Pgn)$+jets
event can be present after applying the lepton veto that is used to select signal events.
The resulting estimate of the $\PW(\to\ell\Pgn)$+jets background is calculated as follows:
\begin{linenomath}
\begin{equation}
\label{eq:ddWJbg}
N(\PW\to \Pgn,\text{lost}\,\ell) = \frac{N^\text{obs}(\Pgm) - N^\mathrm{MC}_\text{non-\PW} }{ A\times\epsilon(\Pgm) }
 \sum_{\ell=\Pe,\,\Pgm,\,\Pgt}~\mathcal{P}(\text{lost}\,\ell)
\end{equation}
\end{linenomath}
where $N^\text{obs}(\Pgm)$ is the observed number of single muon events, $N^\mathrm{MC}_\text{non-\PW}$ is the
background that does not arise from \PW~bosons, and is estimated through simulation,
$A\times\epsilon(\Pgm)$ is the product of acceptance and efficiency to identify and select the muon, as measured in simulation, and
$\mathcal{P}(\text{lost}\,\ell)$ are the probabilities that a \PW+jets event with an electron, a muon, or a tau lepton is not rejected by the signal selection, as defined through simulation.
The background contributions from other kinematic regions [e.g. $\PW(\to\Pe\Pgn)$ and $\PW(\to\Pgt\Pgn)$ with two jets] were found to be negligible.
The accuracy of the $\PW(\to\ell\Pgn)$+jets background estimate is limited by the
number of selected muon events and by the uncertainty in the simulation of background from other than \PW~boson sources.
The rate of \PW+jets events with one \cPqb-tagged jet is estimated by scaling the rate without \cPqb-tagged jets
by the probability to have a \cPqb-tagged jet in simulated \PW+jets events.
The estimated background from \PW+jets is given in Table~\ref{tab:mc2data}.

The most important background after $V$+jets processes is from
\ttbar production, followed by single top quark and diboson production.
These backgrounds are estimated through simulation. The leading systematic
uncertainties arise in the simulated \ttbar sample. They are related to the choice of the renormalization and factorization
scales and the scale that determines the transition between modeling
additional partons at matrix element level and at the level of parton
showers.
Other systematic uncertainties originate from jet energy scale and resolution, \cPqb-tagging efficiency
and mistagging rate, choice of PDF, and accuracy of the luminosity measurement.
The yields from background \ttbar, single top quark, and diboson sources,
together with the systematic uncertainties, are given in Table~\ref{tab:mc2data}.

\begin{figure}[htbp]
\centering
\includegraphics[width=\cmsFigWidth]{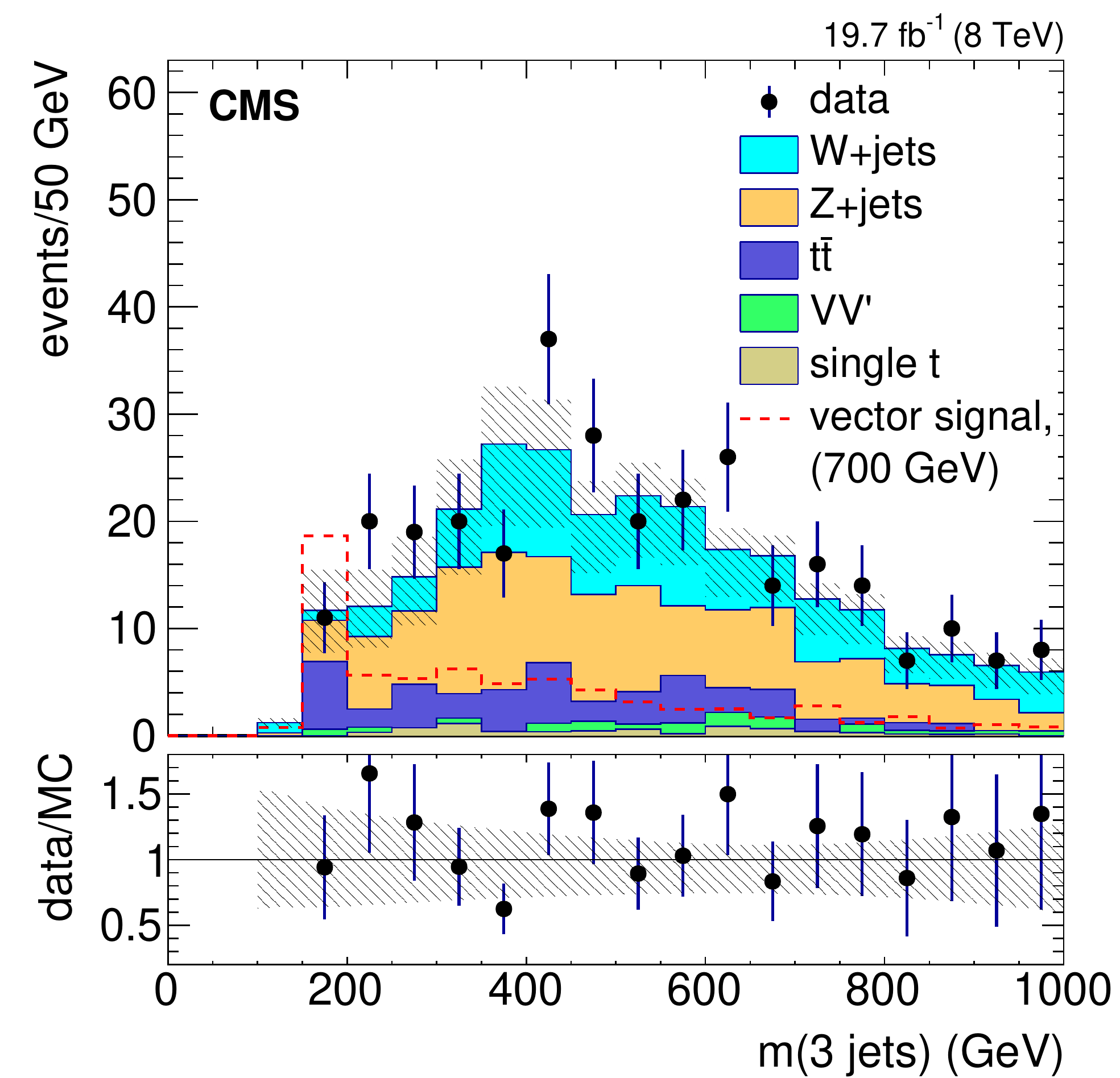}
\caption{The invariant mass of the three jets prior to the selection on their mass to be   less than 250\GeV, for events with one \cPqb-tagged jet.
Data are compared to the simulated backgrounds.
The expectation from a model for an invisible vector particle with a mass of 700\GeV is represented by the dashed line.
\label{fig:triJetMass}}

\end{figure}

Figure~\ref{fig:triJetMass} shows the distribution of the invariant mass of the three jets before
requiring   their invariant mass   to be   less than 250\GeV, in events with one \cPqb-tagged jet.
We do not present a simulation of the multijet background, thus for the comparison
between data and simulated backgrounds we suppress the potential contribution from this source with
an additional cut on the opening azimuthal angle between the two leading jets: $\abs{\phi^\text{jet1}-\phi^\text{jet2}}<2.8$.
The shaded areas represent the sum of the systematic uncertainties related to the renormalization and factorization scales for the
\ttbar and $V$+jets backgrounds, smoothed in a second-order polynomial fit, and taken in quadrature.
Agreement is observed between data and background predictions.
The dashed line in Fig.~\ref{fig:triJetMass} indicates the prediction from a model based on a 700\GeV invisible vector boson.

The signal cross section, as well as the number of multijet background events, are measured in
data using a likelihood approach, where each systematic source is treated as a nuisance parameter.
The method is based on the observed number of events without and with just a single \cPqb-tagged jet
accepted in selecting the signal. These two event categories contain untagged and tagged signal
and background events as shown in the following system of equations:
\begin{linenomath}
\begin{equation}\label{eq:bTagRateMatrix}
\begin{array}{ll}
N^\mathrm{0b} = \mathcal{P}_\text{sig}^\mathrm{0b}\cdot N_\text{sig} + \mathcal{P}^\mathrm{0b}_\mathrm{MJ}\cdot N_\mathrm{MJ} + N^\mathrm{0b}_\mathrm{other}, \\
N^\mathrm{1b} = \mathcal{P}_\text{sig}^\mathrm{1b}\cdot N_\text{sig} + \mathcal{P}^\mathrm{1b}_\mathrm{MJ}\cdot N_\mathrm{MJ} + N^\mathrm{1b}_\mathrm{other},
\end{array}
\end{equation}
\end{linenomath}
where $\mathcal{P}_\text{sig}^\mathrm{0b}$ and $\mathcal{P}_\text{sig}^\mathrm{1b}$ are the probabilities to tag 0 or 1 jet as a \cPqb~jet in the
selected signal events, $\mathcal{P}_\mathrm{MJ}^\mathrm{0b}$ and $\mathcal{P}_\mathrm{MJ}^\mathrm{1b}$ are the corresponding probabilities for
the selected multijet events in data, and $N^\mathrm{0b}_\mathrm{other}$ and $N^\mathrm{1b}_\mathrm{other}$ are the known contributions
to 0 and 1 \cPqb-tagged event categories from other backgrounds.
The $\mathcal{P}_\mathrm{sig,MJ}^\mathrm{0b,1b}$ probabilities are estimated using simulation.
The uncertainty in the $\mathcal{P}_\mathrm{MJ}^\mathrm{0b,1b}$ probabilities is taken as the difference between the estimate
obtained from simulation and that obtained from data using a control region defined by relaxing the \MET requirement.
The system above is solved to estimate the number of multijet ($N_\mathrm{MJ}$) and signal ($N_\text{sig}$) events, by using a numerical minimization of the following likelihood:
\begin{linenomath}
\begin{equation}
\label{eq:likelihood}
\begin{split}
\L(\sigma_\text{sig}, \bnu) =
  \text{Poisson} \left( N_\text{obs}^\mathrm{0b} \big|
  N^\mathrm{0b}
 \right)
\times 
  \text{Poisson} \left( N_\text{obs}^\mathrm{1b} \big|
  N^\mathrm{1b}
 \right),
\end{split}
\end{equation}
\end{linenomath}
where $\sigma_\text{sig}$ is the signal cross section,
$\bnu$ is the vector of the nuisance parameters describing uncertainties in the expected number of events from the Eq.~(\ref{eq:bTagRateMatrix}), and $N_\text{obs}^\mathrm{0b}$ and $N_\text{obs}^\mathrm{1b}$ are, respectively,
the total number of observed events in event categories without and with one \cPqb~tag.

The number of expected SM background events is compared to the data after applying the final selections, and is presented in Table~\ref{tab:mc2data}.
Systematic uncertainties in the simulated backgrounds (\ttbar, single top, and $VV'$) are presented as sums of the
uncertainties from all of the respective sources, taken in quadrature. The multijet background is calculated using all the other backgrounds
and data in Eq.~(\ref{eq:likelihood}). The uncertainty in the multijet background is determined by the uncertainties
in the other backgrounds, and is therefore not included in the quadratic sum of background uncertainties.

\begin{table}[htb]
\centering
\topcaption{Total number of selected events in data compared to the background prediction.
The background yields are given with statistical (first) and systematic (second) uncertainties.
The multijet background is calculated using all the other backgrounds and therefore its
uncertainty is not included in the quadratic sum of background uncertainties.
\label{tab:mc2data}}
{

\begin{scotch}{lcc}
             &    No \cPqb~tag   &   One \cPqb~tag  \\ \hline
\ttbar       &   6$\pm$0$\pm$5   &  12$\pm$0$\pm$12  \\
\PW+jets     &  18$\pm$9$\pm$7   &   3$\pm$1$\pm$2   \\
\cPZ+jets    & 103$\pm$33$\pm$9  &  11$\pm$10$\pm$1  \\
Single top   &   2$\pm$1$\pm$1   &   1$\pm$1$\pm$1   \\
$VV'$    &   5$\pm$0$\pm$0   &   0$\pm$0$\pm$0   \\
Multijet     &     6($\pm$39)    &    1($\pm$9)      \\
Total background   &    140$\pm$36     &    28$\pm$16      \\
Signal       &      2$\pm$6      &     3$\pm$11      \\
Data         &        143        &       30          \\
\end{scotch}
}
\end{table}

No excess is observed above the background expectation, and limits are set at
95\% confidence level (\CL).
The limits are calculated using the \CLs
technique, which is based on statistical inference method jointly adopted by the ATLAS and CMS
collaborations for the Higgs boson searches~\cite{ATLAS:2011tau}.
The resulting limits are calculated using the expected
signal and background predictions along with their
uncertainties, and the likelihood given in Eq.~(\ref{eq:likelihood}).
Statistical uncertainties, arising from number of observed events with
one or two muons in the control regions are modeled with Poisson
probabilities while all other uncertainties are modeled as log-normal
distributions.

\begin{figure}[htbp]
\centering
\includegraphics[width=0.48\textwidth]{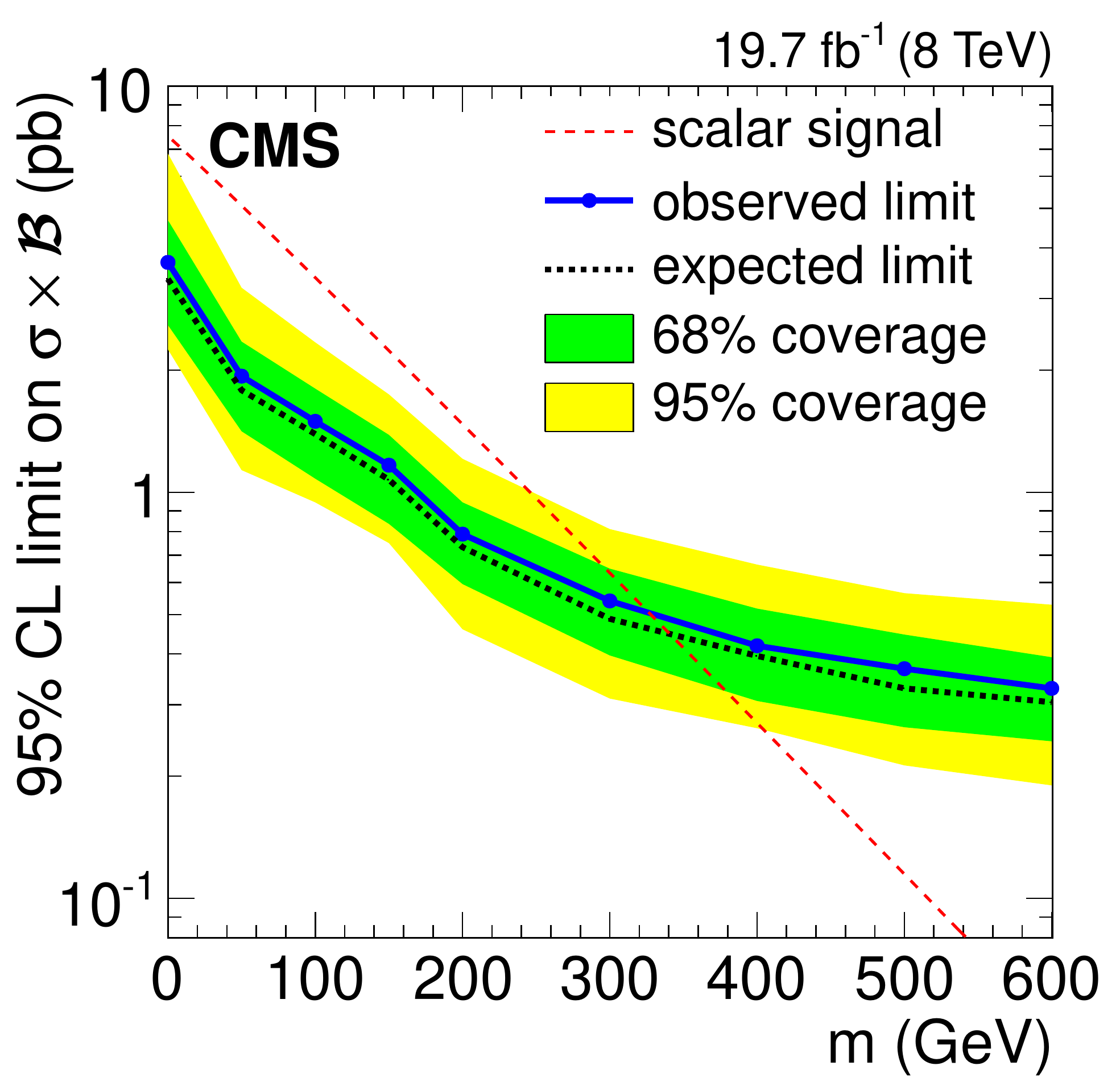}
\includegraphics[width=0.48\textwidth]{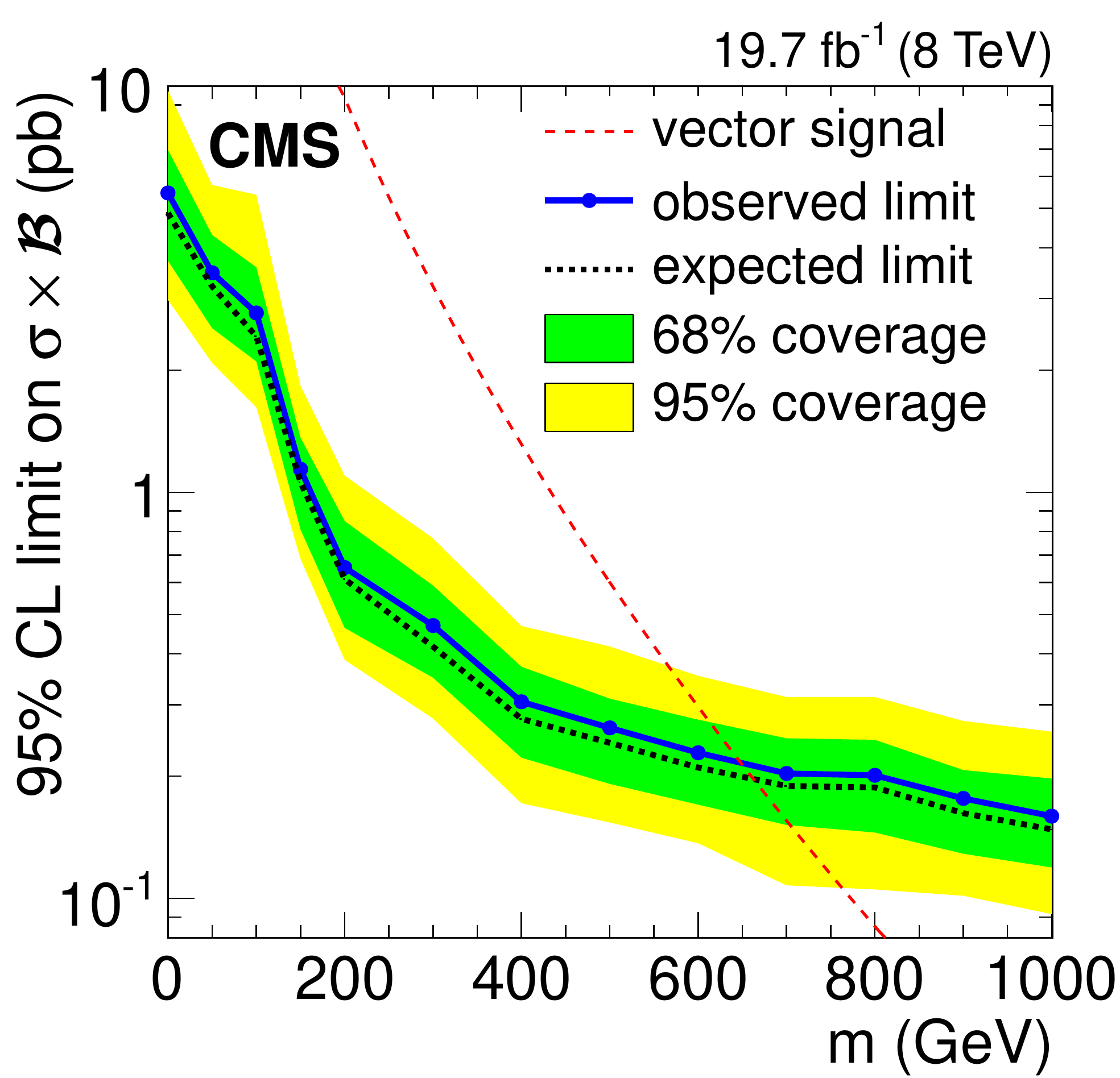}
\caption{The 95\%\,\CL expected and observed \CLs limits as functions of the mass of a
scalar (\cmsLeft) and vector (\cmsRight) invisible particle.
The expected magnitude of a signal as a function of mass, calculated at
leading order, is shown by the dashed curve.
The confidence intervals for the expected limit are given at 68\% and 95\% coverage probability.}
\label{fig:limits}
\end{figure}

Figure~\ref{fig:limits} shows the 95\%\,\CL expected and observed limits
on the product of the production cross section of the monotop and the branching ratio of the \PW~decay
to $\cPq\cPaq'$, as a function of mass of the invisible bosonic state, for scalar and vector fields.

In summary,
a search has been performed by the CMS Collaboration for invisible particles produced in association with a single top quark
that decays into three jets, one of which is \cPqb-tagged.
The results are interpreted using a monotop model that predicts the existence of invisible scalar or vector particles.
The signal and the backgrounds are extracted using a likelihood-based method.
No excess of data over the standard model prediction is found and exclusion limits are set at 95\% confidence level.
The observed lower limits on mass for invisible scalar and vector particles are set at 330\GeV and 650\GeV, respectively.
   For a coupling constant $a_\mathrm{FC}$~=~0.2 these limits increase to 530\GeV and 930\GeV, respectively.
These results substantially extend a previous limit on monotop production of an invisible vector particle published by the CDF Collaboration~\cite{Aaltonen:2012ek}
and complement the 8\TeV results of the ATLAS Collaboration~\cite{Aad:2014wza} obtained with the leptonic top quark decay channel.

\begin{acknowledgments}
We congratulate our colleagues in the CERN accelerator departments for the excellent performance of the LHC and thank the technical and administrative staffs at CERN and at other CMS institutes for their contributions to the success of the CMS effort. In addition, we gratefully acknowledge the computing centers and personnel of the Worldwide LHC Computing Grid for delivering so effectively the computing infrastructure essential to our analyses. Finally, we acknowledge the enduring support for the construction and operation of the LHC and the CMS detector provided by the following funding agencies: BMWFW and FWF (Austria); FNRS and FWO (Belgium); CNPq, CAPES, FAPERJ, and FAPESP (Brazil); MES (Bulgaria); CERN; CAS, MoST, and NSFC (China); COLCIENCIAS (Colombia); MSES and CSF (Croatia); RPF (Cyprus); MoER, ERC IUT and ERDF (Estonia); Academy of Finland, MEC, and HIP (Finland); CEA and CNRS/IN2P3 (France); BMBF, DFG, and HGF (Germany); GSRT (Greece); OTKA and NIH (Hungary); DAE and DST (India); IPM (Iran); SFI (Ireland); INFN (Italy); NRF and WCU (Republic of Korea); LAS (Lithuania); MOE and UM (Malaysia); CINVESTAV, CONACYT, SEP, and UASLP-FAI (Mexico); MBIE (New Zealand); PAEC (Pakistan); MSHE and NSC (Poland); FCT (Portugal); JINR (Dubna); MON, RosAtom, RAS and RFBR (Russia); MESTD (Serbia); SEIDI and CPAN (Spain); Swiss Funding Agencies (Switzerland); MST (Taipei); ThEPCenter, IPST, STAR and NSTDA (Thailand); TUBITAK and TAEK (Turkey); NASU and SFFR (Ukraine); STFC (United Kingdom); DOE and NSF (USA).
\end{acknowledgments}

\bibliography{auto_generated}

\cleardoublepage \appendix\section{The CMS Collaboration \label{app:collab}}\begin{sloppypar}\hyphenpenalty=5000\widowpenalty=500\clubpenalty=5000\input{B2G-12-022-authorlist.tex}\end{sloppypar}
\end{document}

%% file: B2G-12-022-authorlist.tex
\textbf{Yerevan Physics Institute,  Yerevan,  Armenia}\\*[0pt]
V.~Khachatryan, A.M.~Sirunyan, A.~Tumasyan
\vskip\cmsinstskip
\textbf{Institut f\"{u}r Hochenergiephysik der OeAW,  Wien,  Austria}\\*[0pt]
W.~Adam, T.~Bergauer, M.~Dragicevic, J.~Er\"{o}, C.~Fabjan\cmsAuthorMark{1}, M.~Friedl, R.~Fr\"{u}hwirth\cmsAuthorMark{1}, V.M.~Ghete, C.~Hartl, N.~H\"{o}rmann, J.~Hrubec, M.~Jeitler\cmsAuthorMark{1}, W.~Kiesenhofer, V.~Kn\"{u}nz, M.~Krammer\cmsAuthorMark{1}, I.~Kr\"{a}tschmer, D.~Liko, I.~Mikulec, D.~Rabady\cmsAuthorMark{2}, B.~Rahbaran, H.~Rohringer, R.~Sch\"{o}fbeck, J.~Strauss, A.~Taurok, W.~Treberer-Treberspurg, W.~Waltenberger, C.-E.~Wulz\cmsAuthorMark{1}
% ...
\vskip\cmsinstskip
% ...
1:~~Also at Vienna University of Technology, Vienna, Austria\\
2:~~Also at CERN, European Organization for Nuclear Research, Geneva, Switzerland\\
% ...
